\documentclass[final,5p,times,twocolumn]{elsarticle}

\usepackage{lineno}
\usepackage{xcolor}
\usepackage{soul}
\usepackage{multirow}
\usepackage{hyperref}
\usepackage{amsmath}

\journal{Acta Astronautica}

\begin{document}
\begin{frontmatter}
\title{Orbit Design for the Millimetron Space Observatory}
\author[1]{T.~A.~Syachina\corref{cor}}
\author[1]{A.~G.~Rudnitskiy}
\author[1]{P.~V.~Mzhelskiy}
\author[1]{M.~A.~Shchurov}
\author[1]{P.~R.~Zapevalin}
\affiliation[1]{
    organization={Astro Space Center, Lebedev Physical Institute, Russian Academy of Sciences},
    addressline={Profsoyuznaya str. 84/32}, 
    city={Moscow},
    postcode={117997}, 
    country={Russian Federation}}
\cortext[cor]{Corresponding author: syachina@asc.rssi.ru}
\begin{abstract}
Millimetron is a space observatory for millimeter and sub-millimeter observations planned for launch around 2030. The 10-meter diameter space unfolded telescope will be cooled down to 10~K and operated in the vicinity of Lagrange point L2. Mission lifetime is 10 years and it includes astronomical observations in two modes: as a space-ground interferometer and as a single-dish telescope. This paper presents the results of the Millimetron space observatory orbit design that takes into account technical and scientific requirements and constraints. It covers the computation of suitable operational orbits, the selection of an orbit, and the transfer from Earth. Furthermore, scientific objectives are demonstrated through VLBI visibility simulation and image reconstruction. Unlike previous works that used analytical methods, this work employs numerical integration for orbital design. Based on the orbital design methods developed in this work, we calculate the exact dates for departure, halo formation, and trajectory correction. Additionally, we investigate the existence of the short baseline projection and its specific dates for VLBI mode and show the feasibility of scientific objectives through VLBI visibility simulation and image reconstruction.
\end{abstract}


\end{frontmatter}

\section{Introduction}
The Millimetron space observatory is an international project being developed under the leadership of the Astro Space Centre of the Lebedev Physical Institute of the Russian Academy of Sciences (ASC LPI RAS). It is an optical Cassegrain telescope with a 10-meter unfolded parabolic antenna. The observatory has two operation modes. First is a single-dish mode (70~$\mu$m-3~mm wavelength) and second -- a very long baseline radio interferometer (VLBI) mode (0.5-10~mm wavelength)\footnote{More details: \url{https://millimetron.ru/en/general}}. The science program is designed to carry out observations for 10 years and covers such areas as relativistic astrophysics and the study of the close vicinity of supermassive black holes (SMBHs), the study of star formation regions, the search for water and complex chemical compounds in these regions, the study of the early Universe and the observation of distortions in the cosmic microwave background.

The observatory is planned to be launched around 2030. In order for all assigned scientific tasks to be accomplished, the project under development needs to be done very carefully. One of the problems is the correct orbit design, which is strongly influenced by technical conditions. To achieve maximum sensitivity of on-board instruments, the antenna and hardware will be actively cooled to 10 and 4 K, respectively. For maintaining the thermal regime of the observatory it was suggested to place the space telescope in the vicinity of the libration point L2 of the Sun-Earth system at a distance of about 1.5 million kilometres from Earth \cite{Andrianov2021,Likhachev2022,Rudnitskiy2022}. This area offers the most effective thermal stability for the entire telescope, avoiding sharp temperature changes.

It is significant to note that Millimetron will be the first to observe at L2 using VLBI mode. Previously, missions launched to libration point operated solely in the single-dish mode and performed either observations of selected sources or a survey of the entire sky like Planck, GAIA, Herschel, Spektr-RG, and James Webb Telescope \cite{Pilbratt2010, Tauber2010, Mignard2014, Kovalenko2019, Gardner2023}. For the past decades, successful space-based VLBI missions have exclusively been conducted in near-Earth orbits (HALCA (VSOP) and Radioastron missions) \cite{Hirabayashi2000,Kardashev2013}. New concepts for space and ground-space VLBI interferometer missions also consider near-Earth or near-lunar orbits (for example, Event Horizon Imager (EHI), THEZA (TeraHertz Exploration and Zooming-in for Astrophysics), CAPELLA \cite{Hong2014,Zakhvatkin2020,Kudriashov2021,Gurvits2021,Trippe2023, Rudnitskiy2023}). Therefore, the main difference in the orbit design for the Millimetron observatory is the consideration of all the requirements imposed by the VLBI mode. And with these requirements taken into account, the task of searching, selecting and designing an optimal orbit becomes relevant and novel.

VLBI observations aim to obtain detailed images of supermassive black holes M87* and Sgr~A*. The VLBI mode implies simultaneous observation of the same source by several telescopes separated in space. Such observations allow us to achieve incredibly high angular resolution. And the resolution in this case will be determined by the baseline projection, which is imposed by the VLBI mode requirements.

Previous studies have confirmed that in general the most suitable orbital configuration for Millimetron is a halo orbit in the vicinity of the Sun-Earth L2 libration point \cite{Andrianov2021,Likhachev2022,Rudnitskiy2022}. However, these studies were analytical and did not directly relate to the search for the optimal orbit and its design. Moreover, they did not consider all technical constraints. In this paper, we present the orbit design results for the Millimetron space observatory. We provide orbit parameters, describe its computation method based on limitations, analyze orbit maintenance and examine observational capabilities.

This article consists of 5 sections. Section 2 covers all orbit requirements and constraints. Section~3 describes the orbit design and calculation procedure, provides the maintenance and transfer from Earth estimates. Section 4 describes the mission's scientific capabilities for the designed orbit. Through VLBI visibility simulation and image reconstruction of M87 source, we demonstrate the validity of the orbit. Section 4 also provides the results of radio visibility estimates for the Millimetron observatory from ground tracking stations. The final Section 5 summarizes and discusses the results of the performed computation and analysis.

\section{Mission Constraints}\label{sec:orb_conf} 

The orbit design of any space mission must take into account all the requirements and limitations imposed by the scientific program and the scientific problems to be solved. In addition, it must take into account the specific technical constraints for a specific mission.

\subsection{Scientific constraints} \label{sec:sci_cons}
First, let us consider scientific requirements and constraints. As mentioned earlier, the space telescope will observe in two modes. Thus, there are two constraints each related to its own observing mode.

\begin{itemize}
    \item Single-dish mode requirement: annual access to the entire sky.
\end{itemize}

\begin{figure}[h]
    \centering
    \includegraphics[width=0.8\linewidth]{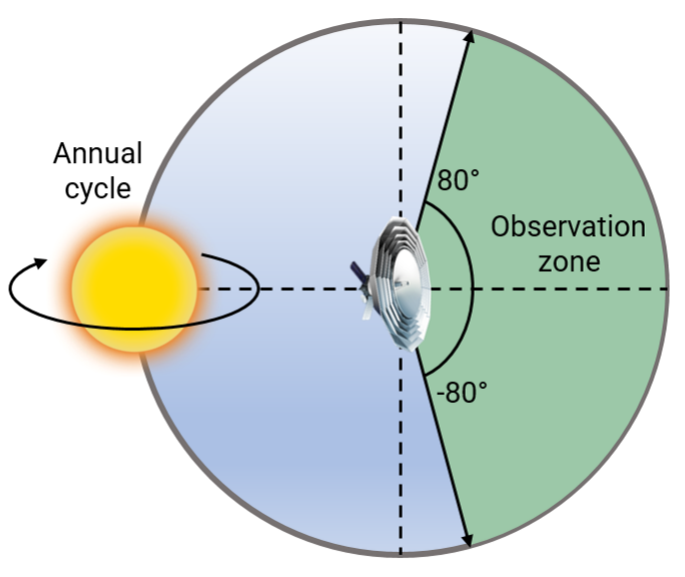}
    \caption{The observatory at libration point L2 has the ability to observe the sky in a range of $\pm$ 80 degrees relative to the ecliptic plane.}
    \label{fig:obs}
\end{figure}

This requirement is satisfied by the orbit type itself. Halo orbit in the vicinity of the Sun-Earth L2 libration point has several advantages, such as the ability to observe the entire sky and the absence of illumination from celestial bodies. The field of view is limited to an area of 160$^{\circ}$ relative to the ecliptic plane (Fig.~\ref{fig:obs}). The estimated sky coverage map can be seen in Fig.~\ref{fig:mol}.
  
\begin{itemize}
    \item Space-ground VLBI mode requirement:\\minimal space-ground baseline projections.
\end{itemize}  

The main parameters of a VLBI interferometer that fundamentally affect the reconstructed image are sensitivity and angular resolution. The angular resolution $\theta$ is expressed as follows:

\begin{equation}
    \theta \sim \frac{\lambda}{D},
\end{equation}
where $\lambda$ -- observing wavelength, $D$ -- baseline projection of two antennas of VLBI interferometer.

\begin{figure}[h]
    \centering
    \includegraphics[width=0.8\linewidth]{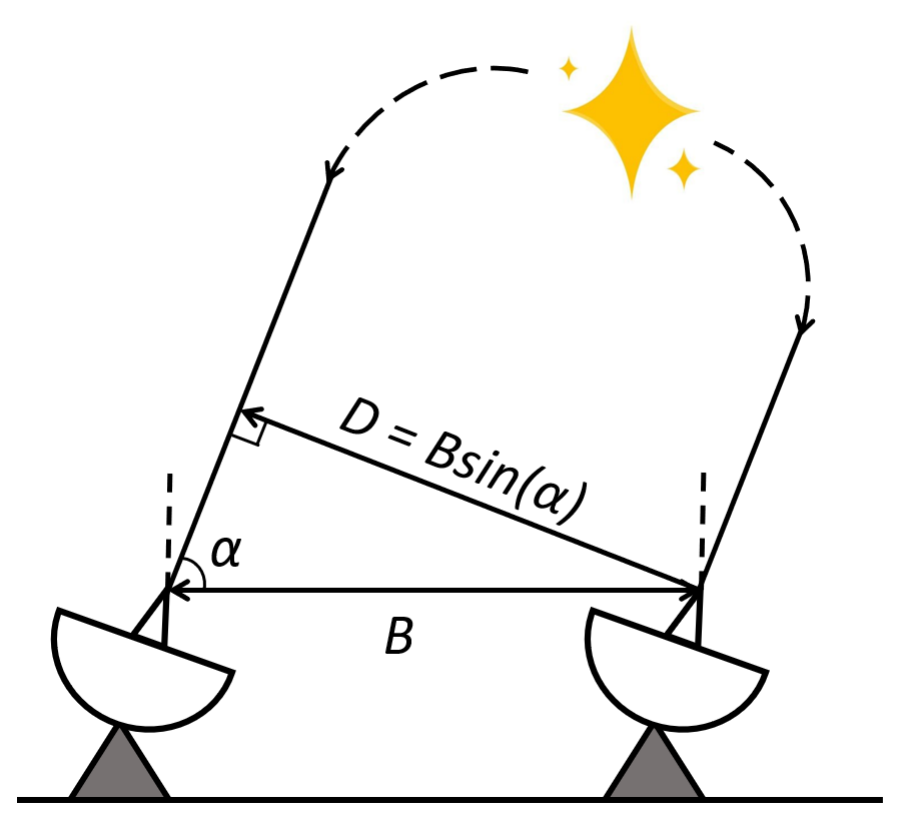}
    \caption{Schematic geometric configuration of a single baseline VLBI interferometer. $B$ -- a distance between the two radio telescopes (baseline), $\alpha$ -- angle between the baseline $B$ and the direction to the observing source, $D$ - baseline projection.}
    \label{fig:bl_vlbi}
\end{figure}

Baseline projection is determined as
\begin{equation}
    D =B\sin{\alpha},
\end{equation}
where $B$ -- distance between the two radio telescopes (baseline), $\alpha$ -- angle between the baseline $B$ and the direction to the target source. So, $D$ -- is a distance between two telescopes, projected (baseline projection) onto the sky plane in a source direction or $(u,v)$-plane. An example of a single baseline VLBI interferometer is shown in Fig.~\ref{fig:bl_vlbi}.

A condition is introduced: orbit must provide minimal ($\sim$0.8~-0.9~ED (Earth diameters)) baseline projections together with ground telescopes for a number of selected sources (primarily Sgr~A* and M87) to perform VLBI imaging experiments. Such values of the baseline projection are assured to obtain the intersection of space-ground and purely ground baselines for two-dimensional imaging of SMBHs in VLBI experiments. This is based on the experience of the Radioastron observatory \cite{Kardashev2013}. However, in the case of a halo orbit in L2, the search for such regions for two or more astronomical sources is a complex task. This is due to slow orbital evolution.

\subsection{Technical constraints}

\begin{figure}[h]
    \centering
    \includegraphics[width=0.8\linewidth]{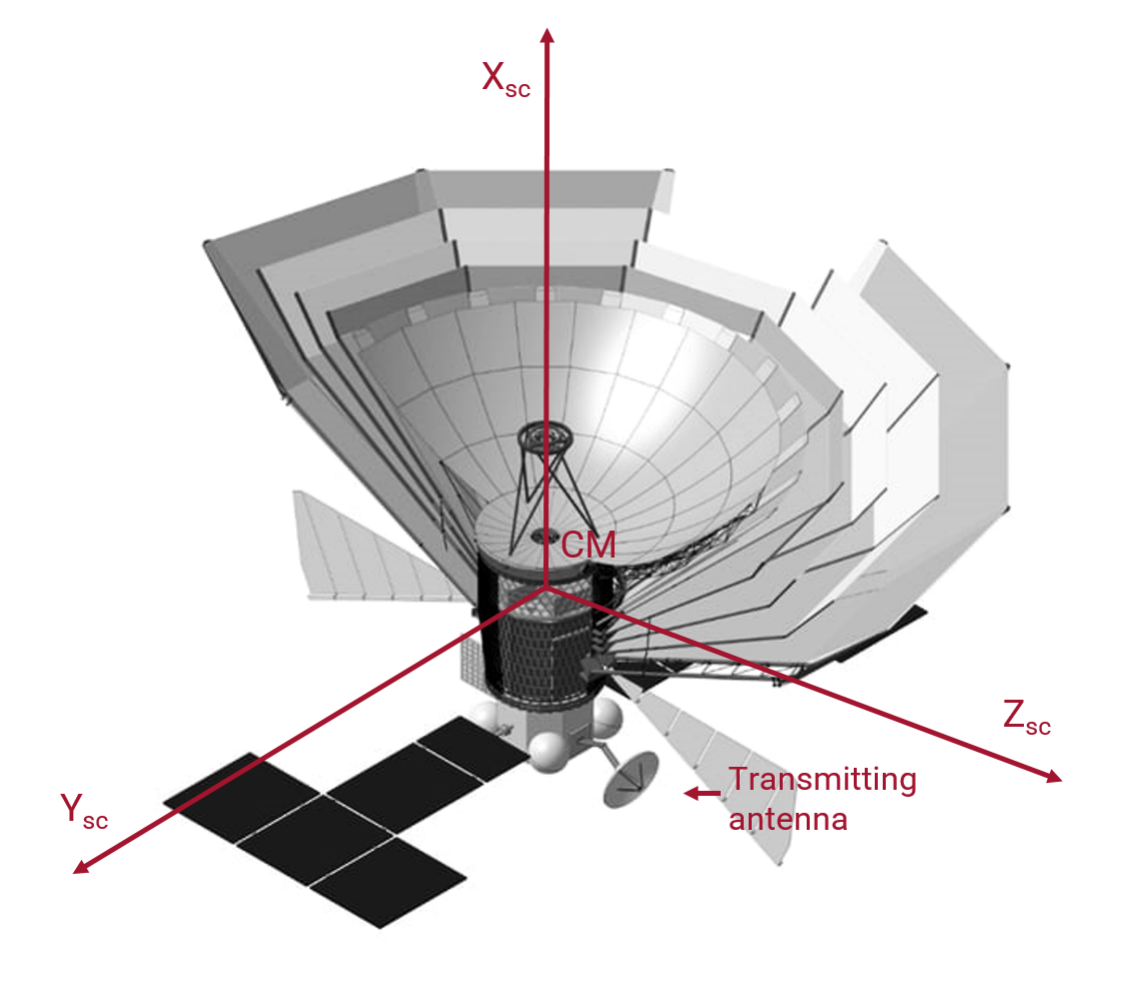}
    \caption{The coordinate system of the spacecraft. The origin is located at the point of the center of mass. The $X_{sc}$ axis is directed along the center axis of the spacecraft. $Y_{sc}$ axis is directed along the solar panels and $Z_{sc}$ axis completes the system to the right. Additionally, the transmitting antenna is indicated in the figure.}
    \label{fig:sc_axes}
\end{figure}

The schematic view of the spacecraft is shown in Fig.~\ref{fig:sc_axes}. The main mirror diameter is 10 meters, which is too large for the existing launch vehicle fairings, so the mirror structure will be unfolded. The basic module is “Navigator-SM”. It provides control of the spacecraft and guidance of the onboard complex of scientific equipment for astronomical objects. It also solves the electrical needs of the spacecraft. The module is equipped with a high-gain antenna for data transmission to Earth. The observatory will be launched on an Angara-5M launch vehicle with ``DM3 Stage 2” as the upper stage. The mass of the entire spacecraft will be 6,600~kg, including 800~kg of fuel. The spacecraft is expected to carry correction engines with a specific impulse of 284.229~s and a thrust of 53.9~N.

Most technical requirements are defined by the observatory's thermal regime. Both single-dish and VLBI modes require active onboard cooling to provide necessary operating temperatures for the sensitive scientific onboard instruments. The antenna must be cooled to 10~K and the instruments to 4~K. Therefore, during mission operation, a constant orientation must be maintained. The telescope will always be directed anti-solar, with the spacecraft's back facing the Sun. In this paper, orientation correction problems are not explored. Due to the selected orbit type near the L2 point of the Sun-Earth+Moon system, such requirements are relatively easy to obtain. Even when the spacecraft is launched into an intermediate orbit around Earth, the main mirror will not be illuminated as the angle between the velocity vector and the Sun relative to the spacecraft will be $153^\circ$.

Other constraints are mostly related to data transmission and each limits orbital parameters:

\begin{itemize}

    \item Absence of penumbral shadows from the Moon and the Earth to ensure stable power supply from solar batteries to the onboard equipment. The shadow cone radius at the L2 point distance is $\approx$ 13 000 km. For 10 years of the planned scientific program, the observatory should stay out of the exclusion zone.
    \item The angle between the Sun, the satellite and the ground tracking station must be $\alpha \geq 10^{\circ}$ to avoid possible damage to the transmitting antenna. 
    \item It is essential that at least one ground station has permanent radio visibility of the spacecraft. Four possible ground tracking stations are considered for the Millimetron spacecraft data downlink and orbit determination: Baikonur, Ussuriysk, Bear Lakes and Evpatoria.
    \item VLBI observations should be conducted no sooner than one month after entering orbit to set up the spacecraft and ensure that all systems are operational.
    \item VLBI observations of M87 and Sgr~A* should be made within two days. Consequently, this requirement arises because of the limited capacity of the onboard memory (384~Tb).
 \end{itemize} 

\section{Orbit Design}
\label{sec:orbit}

We divide the orbit design task into three problems: computation of the halo orbit in the vicinity of the libration point L2 in the Sun-Earth+Moon system, a transfer from Earth to L2, and maintaining this orbit during the spacecraft's lifetime.

This section discusses an influence of constraints on orbit parameters. It is known that the halo orbit can be described by a single parameter, the Jacobi constant or the altitude above the ecliptic plane $A_z$. For better clarity the amplitude is used as a parameter. Furthermore, halo orbits are divided into northern and southern depending on the rotation direction. Based on the conditions under consideration orbits within $320\times10^{3}$~km $<A_z< 445\times10^{3}$~km should be considered to provide suitable radio visibility. Compliance with other conditions will be discussed below.

\subsection{Orbit Computation}

A significant difference in orbit design, compared to other missions, is the need to select an orbit that meets VLBI mode observatory requirements. This approach is being implemented for the first time, since the Millimetron observatory will be the first observatory to perform VLBI observations at the L2 point.

The process of searching for an orbit involves several stages. First, the analytical approximation in the system of the circular restricted three-body problem (CR3BP) is calculated. The well known equations of motion of a third massless body in this system can be written as

\begin{equation}
\begin{aligned}
  \ddot{x}-2\dot{y} = -U_x, \\
  \ddot{y}+2\dot{x} = -U_y, \\
  \ddot{z} = -U_z.  
\end{aligned}
\label{eqmotion}
\end{equation}

The distances, velocities and time are normalized by the distance between the primaries, their angular velocity and orbital period, respectively. The effective potential $U$ is:

\begin{equation}
    U = -\frac{1}{2}(x^2+y^2) - \frac{1-\mu}{r_1} - \frac{\mu}{r_2},
\label{effpotential}
\end{equation}

where $\mu = m_2/(m_1+m_2)$ and ${m_1, r_1}$, ${m_2, r_2}$ correspond to the masses and distances of the spacecraft from primary and secondary bodies. For accuracy the third order approximation of motion equations obtained by Richardson is used \cite{Richardson}. Afterwards, taking advantage of the orbit's symmetry about the $xz$ plane, the analytical approximation is refined numerically for a closed trajectory formation. The initial state vector of the orbit is located at the point with maximum altitude above the ecliptic plane $A_z$, where the following condition is satisfied:
\begin{equation}
    y = \dot{x} = \dot{z} = 0
\label{symcondition}
\end{equation}
After this step it can be determined whether the orbit is suitable for the scientific goals of the VLBI mode. Thus the condition of existence of minimal baseline projections for the two target sources M87 and Sgr~A* is checked.

As it was mentioned, the main parameters of a VLBI interferometer that fundamentally affect the reconstructed image are sensitivity and angular resolution. However, to obtain a relatively high fidelity image of SMBH it is necessary to have a dense $(u,v)$-coverage \cite{Thompson2017}. In this case, we can describe the task of finding $(u,v)$-coordinates on this plane using the following relation:

\begin{equation}
    \begin{pmatrix}
    u \\
    v \\
    w \\
    \end{pmatrix} = \frac{1}{\lambda}\cdot
    \begin{pmatrix}
        \sin{H} & \cos{H} & 0\\
        -\sin{\delta}\cdot \cos{H} & \sin{\delta}\cdot\sin{H} & \cos{\delta}\\
        \cos{\delta}\cdot \cos{H} & -\cos{\delta} & \sin{\delta}\\
    \end{pmatrix}
    \cdot
    \begin{pmatrix}
        L_{x}\\
        L_{y}\\
        L_{z}\\
    \end{pmatrix}
    \label{eq:uv}
\end{equation}

where $\lambda$ -- observing wavelength, $\delta$ -- declination, $H$ -- the angle of Earth's rotation, which is related to the observation time, $L_{x,y,z}$ -- baseline vector coordinates.
And the baseline projection can be expressed through $(u,v)$ coordinates as $D=\sqrt{u^2+v^2}$. Thus, to get a high resolution, a condition on the baseline $D$ must be imposed.
In turn, $D$ is calculated as 
\begin{equation}
\begin{aligned}
  D = r_{sc} \cdot \sin{\angle(r_{sc},r_{src})}, 
\end{aligned}
\label{baseline}
\end{equation}
where $r_{sc}$ is radius vector of spacecraft and $\angle(r_{sc},r_{src})$ is an angle between the direction to the spacecraft and to the source in J2000 coordinate system. The target orbit must satisfy the following condition:
\begin{equation}
\begin{aligned}
  0.8 ED < \min{D}< 0.9
\end{aligned}
\label{baselinecond}
\end{equation}

For baseline searching, three parameters of halo orbits are varied:
\begin{enumerate}
\item Altitude $A_z$ above the ecliptic plane: from $320\times10^{3}$~km to $445\times10^{3}$~km. Radio visibility is achieved at these altitudes.
\item Direction of orbit: northern or southern.
\item Since the launch will take place in 2030, the initial state vector is searched each day during this year.
\end{enumerate}

The orbit is integrated over two loops in the CR3BP system. Then the state vector of each point is translated into the J2000 coordinate system, and the baseline projection at each point is calculated. The minimum baseline constraint indicates that the direction of the source must be close to the direction of the spacecraft relative to the ground. Considering M87 and Sgr~A* as target sources, it was found that only southern halo orbits are suitable for our purposes. The northern halo orbits did not show appropriate baseline projections, but quasi-halo orbits can be considered for further studies. By testing the orbits on different scales, the southern orbit with an altitude of $370\times10^{3}$~km above the ecliptic plane was chosen. The most suitable start dates were in the range from 2030-09-09 to 2030-09-15. Minimal baseline projections satisfy condition (Eq.~\ref{baselinecond}) for both sources in this case: $D\approx11000$~km.

Following the extraction of an orbit with a different phase, the next step is to determine the precise start date of the orbit. The obtained orbit is transferred into a force model using differential correction for each day from 2030-09-09 to 2030-09-15 and a suitable date for condition (Eq.~\ref{baselinecond}) is determined. For the purpose of transferring into a force model one initial vector is integrated into the J2000 coordinate system. Afterwards, the state vectors of the 10~points on the trajectory (every 45~days) are varied to be close to the analytical approximation. Thus two stable halo orbit loops can be provided. Force model includes both central field and non-central harmonics up to order 32~according to the EGM(96)\footnote{EGM(96) model description: \url{https://www.usna.edu/Users/oceano/pguth/md_help/html/egm96.htm}}, as well as the contribution of the solar system bodies: the Sun, all planets, Pluto, and the Moon using the JPL ephemeris\footnote{JPL ephemeris description: \url{https://ssd.jpl.nasa.gov/planets/eph_export.html}}. Runge-Kutta integration method 7(8) is used in all calculations.

\begin{figure*}
\centering
\includegraphics[width=\textwidth]{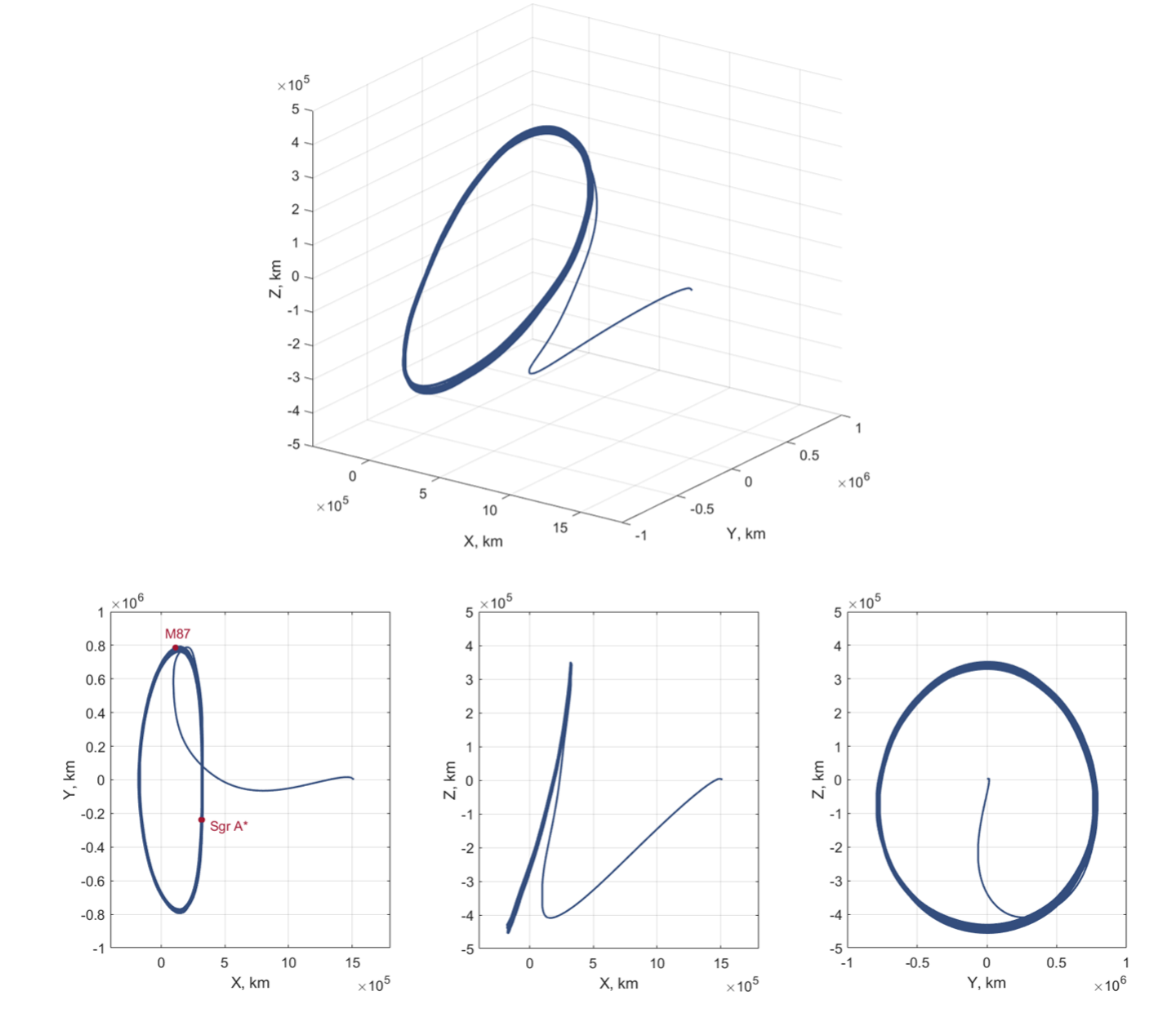}
\caption{Halo orbit and projections for 10 years in L2-centric coordinate system. The observation moments of M87 and Sgr A* are shown by red dots.}
\label{fig:projections}
\end{figure*}

As a result, 2030-09-11 was chosen as a date for starting orbit around L2. This date shows the most suitable baseline projections: $D=11637$~km for M87 and $D=10827$~km for Sgr~A*. Fig.~\ref{fig:projections} illustrates the integrated halo orbit around Lagrange L2 point of the Sun-Earth system in L2-centric inertial coordinate system with coordinate origin at L2 point. On the upper image, we can see the trajectory in 3D and on the bottom, we can see projections from left to right: XY, XZ, YZ. The red dots in the figure show the sections of the trajectory where VLBI imaging observations will be carried out for M87 and Sgr~A*. The altitude from the ecliptic plane $A_z$ is about $330\times10^3$~km to the north and $430\times10^3$~km to the south. The orbital period is 178 days. Total maintenance impulse is $\Delta$v=10.713~m/s for 10 years, which corresponds to $\sim$26~kg of fuel.

\begin{figure}[h]
    \centering
    \includegraphics[width=0.8\linewidth]{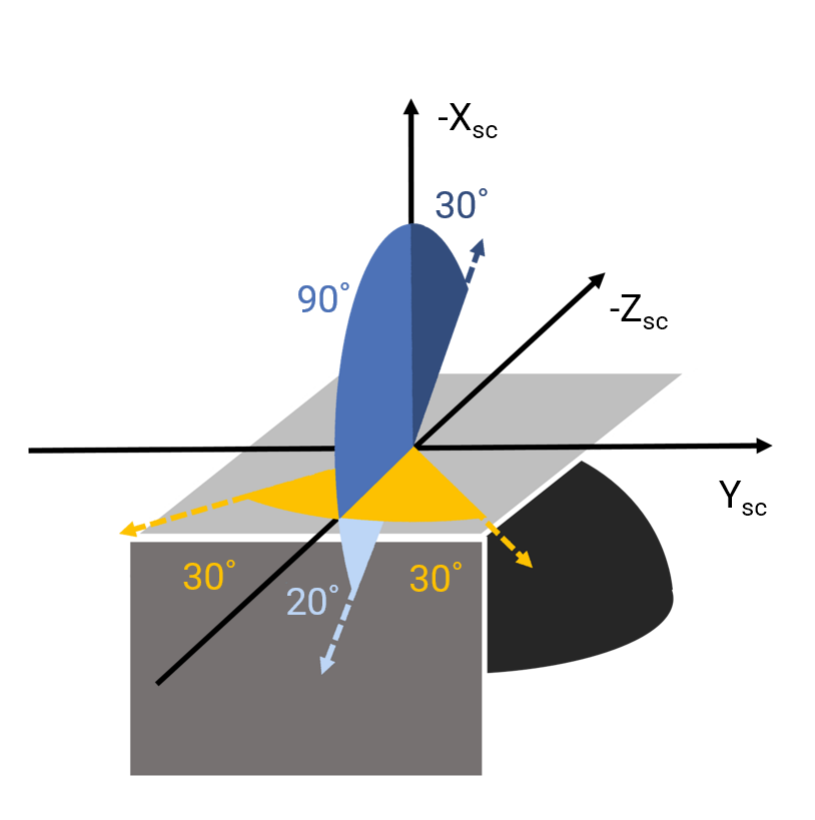}
    \caption{Possible positions of transmitting antenna.}
    \label{fig:angle30}
\end{figure}

Because of its size, the orbit does not fall within the Earth's shadow at all operational times. The angle between the Sun, the satellite and the tracking station is $15^{\circ} < \alpha < 30^{\circ}$, which results in the possible location of a transmitting antenna as shown in Figure~\ref{baseline}. The swing angles of the transmitting antenna are $140^{\circ}$ in the XZ plane of the telescope's coordinate system and $60^{\circ}$ in the YZ plane as illustrated in Fig.~\ref{fig:angle30}. Radio visibility and memory capabilities are discussed in the following paragraphs.

\subsection{Orbit Maintaining}

The halo orbit in the vicinity of point L2 is quasi-periodic, and for the stable existence of the satellite there, it is necessary to perform maintenance corrections. Necessary impulses were calculated for 10 years of space observatory existence using gradient descent. This method relies on spatial restrictions introduced during calculations. These restrictions are the maximum and minimum coordinate values, which the spacecraft should not exceed. Orbits with applied correction impulses is computed gradually reducing the value of an impulse. Afterwards, it is checked whether some of the applied impulses led to an increase in orbit duration within the boundaries. As soon as such a pair of ``orbit+impulse" is detected, corrections are made with this recently found impulse.

The frequency of searching for such impulses should be approximately every 90~days. If the correction impulse was less than 5~cm/sec, then such a correction could not be made. This was due to the spacecraft correction engines' accuracy. The correction impulses found for the obtained orbit are presented in the Table~\ref{tab:corrections}.

\begin{table}[ht]
\caption{Orbital \textbf{maintenance} for 10~years. Total~$\Delta$v$\approx$10.713 m/s. \label{tab:corrections}}
\centering
\begin{tabular}{|l|r|}
\hline
Date (at 12:00:00.0 UTC)         & $\Delta $v, (m/s) \\ \hline
2030-11-14                  & 1.284             \\ \hline
2031-02-22                  & 0.229             \\ \hline
2031-06-02                  & 0.117             \\ \hline
2031-10-30                  & 0.264             \\ \hline
2032-02-07                  & 0.142             \\ \hline
2032-05-17                  & 0.259             \\ \hline
2032-08-25                  & 0.210             \\ \hline
2033-03-13                  & 0.557             \\ \hline
2033-09-29                  & 0.420             \\ \hline
2034-04-17                  & 0.537             \\ \hline
2034-09-14                  & 0.503             \\ \hline
2035-04-02                  & 0.459             \\ \hline
2035-10-19                  & 0.610             \\ \hline
2036-03-17                  & 0.776             \\ \hline
2036-10-03                  & 0.537             \\ \hline
2037-01-11                  & 0.151             \\ \hline
2037-06-10                  & 0.635             \\ \hline
2037-09-18                  & 0.102             \\ \hline
2038-04-06                  & 0.664             \\ \hline
2038-09-03                  & 0.200             \\ \hline
2039-03-22                  & 0.527             \\ \hline
2039-06-30                  & 0.210             \\ \hline
2039-10-08                  & 0.107             \\ \hline
2040-03-06                  & 0.415             \\ \hline
2040-08-03                  & 0.381             \\ \hline
2041-02-19                  & 0.415             \\ \hline
Total $\Delta$v (10 years) & 10.713            \\ \hline
\end{tabular}
\end{table}

\subsection{Start Windows and Transfer Estimations}
The Millimetron observatory can be launched from the Baikonur or Vostochny cosmodromes. Both ensure a minimum inclination of the reference circular orbit of 51.7$^{\circ}$, which is due to the incidence areas of the launch vehicle stages and the nose fairing. The reference circular orbit has an altitude of 200~km above the Earth. 

To compute a transfer trajectory from the Earth to the halo orbit in L2, the differential correction method is used. The calculation occurs in several stages as follows:

1. Calculation of the monodromy matrix. The eigenvector of this matrix, corresponding to eigenvalue $<$1, is responsible for orbit formation.

2. Eigenvectors can be determined at any orbit point. By integrating all eigenvectors back in time an invariant manifold is obtained. It serves as a first approximation for finding the transfer path to Earth in a real force model. Among all points, several are selected that pass close to the ecliptic plane and the Earth.

3. The trajectory is calculated in a real force model, implementing two-point shooting. The initial speed of the selected point in halo orbit is changed so that, integrating the state vector to perigee, the trajectory arrives at a given altitude 200~km above the Earth and inclination 51.7$^{\circ}$. 

\begin{table*}[ht]
\centering
\begin{tabular}{|l|c|r|}
\hline
Departure date           & Date of the halo orbit formation & Halo orbit formation impulse, (m/s) \\
\hline
2030-05-18T05:34:32.12   & 2030-09-24T11:30:2.67            & 12.41 \\
\hline
2030-05-18T07:16:11.23   & 2030-09-22T19:00:2.67            & 11.24 \\
\hline
2030-05-18T22:31:53.45   & 2030-09-25T00:40:2.67            & 7.87  \\
\hline
2030-05-31T16:21:15.72   & 2030-10-03T04:20:2.67            & 22.96 \\
\hline
2030-06-01T06:55:41.32   & 2030-10-16T16:50:2.67            & 22.87 \\                         
\hline
\end{tabular}
\caption{Start windows and halo orbit formation impulse.}
\label{tab:st_wnd}
\end{table*}

Table~\ref{tab:st_wnd} presents the calculation results. Departure date is the date of impulse application by the upper stage from the reference circular orbit at an altitude of 200~km from Earth. The date of formation of the halo orbit is the date of the spacecraft's arrival in the vicinity of the libration point L2 of the Sun-Earth system. The third column contains data on the impulse to form a halo orbit. The data is not consistent over time, because it depends on process convergence. The table shows the most optimal solution of 7.87~m/s, and the possible deviation from it.
The launch time from the intermediate orbit is 2030-05-18T22:31:53.5. Time of arrival at the L2 point orbit is 2030-09-25T00:40:02.6.  Optimal baseline projections for M87 are achieved on 2031-03-15 and for Sgr A* on 2031-07-20. Therefore there will be enough time to configure the equipment.

\section{Science Program Feasibility}
Initially, during the orbit calculation process, all technical and scientific limitations were taken into account. However, it is necessary to check the feasibility of the stated scientific tasks for the calculated trajectory. For the single-dish mode, the coverage of the celestial sphere by the space observatory was calculated. In turn, for the VLBI mode, the imaging capabilities of two priority sources M87 and Sgr~A* were assessed.

\subsection{Single-dish Mode}
\begin{figure*}
    \centering
    \includegraphics[width=0.8\linewidth]{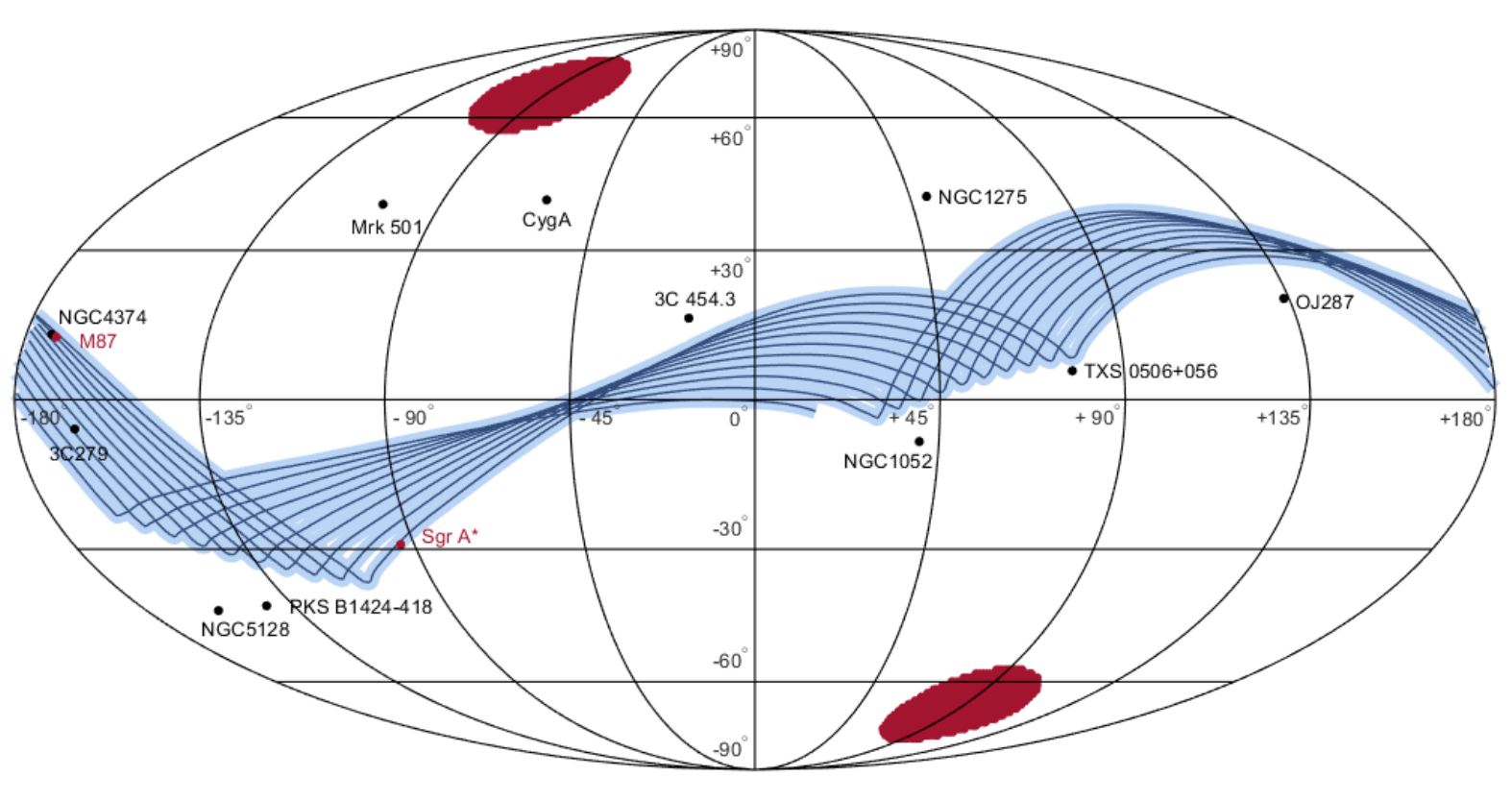}
    \caption{Mollweide projection of the Millimetron observatory trajectory along the calculated orbit for 10 years in the GCRS coordinate system. Light blue line shows the range of baseline projections $\pm$ 5 Earth diameters. Sources from the Table~\ref{tab:target_sources} are plotted with black and red dots. Red zones correspond to areas not available for observations.}
    \label{fig:mol}
\end{figure*}

Observing in a single-dish mode, the main factor determining the possibility of observing a source throughout the year is its location on the celestial sphere. Due to the design features of the observatory, namely the need to avoid illumination of the main mirror by solar radiation, a telescope in halo orbit around the Lagrange point L2 of the Sun-Earth system will be able to observe one orbital revolution, i.e. in about one year. The telescope's viewing area is 0-80$^{\circ}$ degrees from the telescope's primary optical axis. To evaluate observational capabilities in single-dish mode, celestial sphere coverage was assessed for one and 10 years. The results of calculations are shown in Fig.~\ref{fig:mol}. 

It can be seen that in one year Millimetron observes almost the entire celestial sphere. However, it excludes small areas near the ecliptic poles, so $\sim$98\% coverage of the celestial sphere is achieved. This result is maintained for 10 years. In addition, calculations showed that the spacecraft would avoid a penumbral shadow throughout the entire orbital period.

\subsection{Interferometer Mode}\label{sec:VLBIsim}
If for the single-dish mode the assessment of observational efficiency is quite clear, then for the VLBI mode there are certain issues. VLBI mode observes extragalactic sources -- quasars and active galactic nuclei. Moreover one of the key scientific tasks is to obtain highly detailed images of two supermassive black holes Sgr~A* and M87. To perform such observations, it is necessary to implement certain baseline projections of the space-ground interferometer. This is reflected in the list of requirements and restrictions in Section~\ref{sec:orb_conf}.

Simulations were performed to evaluate the observational efficiency of the designed orbit in VLBI mode. It is worth mentioning the results and studies previously conducted in this field \cite{Andrianov2021, Likhachev2022, Rudnitskiy2022}. However, in these works other orbital configurations were considered that were obtained only analytically and did not account for Millimetron observatory constraints. 

Let us consider in more detail the list of sources proposed for VLBI observations. The full list of sources proposed by the Millimetron space observatory scientific group is presented below in Table~\ref{tab:target_sources}. A source was considered observable if its height above the horizon at ground telescopes exceeded $7^{\circ}$. The height was calculated using the World System of Geodetic Parameters WGS-84.

\begin{table*}
\caption{List of target sources for Millimetron Space-VLBI mode.}
\centering
\label{tab:target_sources}
\begin{tabular}{|l|l|l|r|r|r|r|}
\hline
Source   &RA   &Dec		&F$_{\nu}$ at 1.3mm, (Jy)  &$M,~10^{9}\cdot M_{\odot}$   &D, (Mpc)  &$r_g,~(\mu$as)\\
\hline
Sgr~A*   & 17 45 40.0 & $-$29 00 28.2 & 3.5      &$\simeq0.04$ &$\simeq0.008$ & $\simeq 5$\\ 
M87      & 12 30 49.4 & $+$12 23 28.0 & $\sim$ 1 &$\simeq$ 6.5 & 16.8         & $\simeq$4 \\ 
\hline
M84 (NGC4374)  & 12 25 03.7 & $+$12 53 13.0 & $\sim$ 1 &$\simeq$ 1.5 & 16.83        & $\simeq 1$\\ 
3C84 (NGC1275) & 03 19 48.1 & $+$41 30 42.0 & $\simeq$ 10 &$\sim$ 1     & $\simeq$ 73     & $\simeq$ 0.15 \\
OJ287    & 08 54 48.8 & $+$20 06 30.6 & $\simeq3$   & 18          & 930             &
0.22          \\ 
NGS5128 (Cen~A)       & 13 25 27.6 & $-$43 10 08.8 & 5.8         & $0.55-3$    &$\simeq 4$       &
$\simeq 3$    \\
Cyg~A                  & 19 59 28.3 & $+$40 44 02.0 & 1.0         & 2.5     & 232 & 0.11          \\
NGC1052               & 02 41 04.7 & $-$08 15 20.7 &$\simeq$0.5  & 1.5   & 19.3  & 0.05 \\
3C 454.3  & 22 53 57.7 & $+$16 08 53.6 & 1.0         & 40          & 2300 & 0.17 \\   
3C279        & 12 56 11.2 & $-$05 47 21.5 &$\simeq$10   &$\simeq$1    & $z\simeq0.54$   & 0.02          \\
Mrk 501      & 16 53 52.2 & $+$39 45 37.0 &$\simeq$1    &0.5          & 119             & 0.004         \\
TXS 0506+056 & 05 09 25.9 & $+$05 41 35.3 & No data     &0.3          & $z\simeq0.34$   & 0.002         \\
PKS B1424-418&      14 27 56.3      &     -42 06 19.4          & $\sim$ 1 &    No data         & $z=1.52$        &No data\\ 
\hline 
\end{tabular}
\\
$M_{\odot}$ -- Solar mass;
\\
RA -- right ascension in format of hh:mm:ss.ss, where hh -- angular hours, mm -- angular minutes, ss.ss -- angular seconds;
\\
Dec -- declination in format of dd:mm:ss.ss, where dd -- degrees, mm -- angular minutes, ss.ss -- angular seconds.
\end{table*}

All of the sources listed above are supermassive black holes (SMBHs), which are among the closest and brightest objects observed by the Millimetron in VLBI mode. The first two sources, M87 and Sgr~A*, are considered priority sources for high-resolution VLBI imaging. Thus, in both the design of the orbit and in subsequent evaluations of its efficiency in VLBI mode, the problem of obtaining images of these two sources was considered at the outset.

Fig.~\ref{fig:mol} illustrates the Mollweide projection of the Millimetron observatory trajectory along the calculated orbit. Calculations are given for 10 years of observatory active operation. The blue line shows the observatory's trajectory, the cyan line shows the range of space-ground interferometer baseline projections $\sim\pm$5 of Earth diameters. Since the orbit was initially optimized for the imaging task of M87 and Sgr~A*, it can be seen from the figure that the trajectories pass primarily through these two priority sources, thereby providing the necessary short baseline projections. It is worth noting that this still allows imaging of other sources, such as M84 (NGC~4374), which is not that far from M87, and 3C279.

Imaging astronomical sources with VLBI involves a totally different approach than, for example, an optical telescope. A significant feature here is $(u,v)$ coverage. A VLBI technique measures visibility (its amplitude and phase), which is a discrete Fourier coefficients of an astronomical object radio intensity distribution. Each point on the $(u,v)$ plane corresponds to the measured visibility function measurement. The overall distribution of these points on the plane determines image quality or fidelity. Thus, to obtain the radio image of the source it is necessary to obtain $(u,v)$ coverage with measured visibility and perform an inverse Fourier transform of this coverage. As we previously mentioned, it is necessary to implement small baseline projections that intersect with the ground baseline projections of the base. The distribution of points on $(u,v)$ in turn should be as homogeneous as possible.

Simulations consist of calculating the $(u,v)$~coverage, described by~Eq.~\ref{eq:uv} for selected target sources and further image synthesis. The simulation of observations was performed for 230~GHz. In orbit, sections were searched for with small projections of the base and a duration of no more than three days. The duration of a single observation time interval (scan) was chosen to be $t=10$~s, which is due to limitations on coherent integration time on ground telescopes associated with atmospheric fluctuations. The bandwidth of $\Delta\nu=4$~GHz was chosen to match the Millimetron observatory's technical parameters. Also, when calculating $(u,v)$~coverage for M87 and Sgr~A*, limitations of the spacecraft's onboard memory (384~Tb) were taken into account, which directly affect the total duration of observations and the number of scans in them. Two sets of telescopes and their sensitivity and coordinate parameters were considered as ground support: the current Event Horizon Telescope configuration \cite{EHT2019p2} and the extended ngEHT configuration, which includes the current EHT configuration and new promising telescopes \cite{ngEHT2018}.

\begin{figure*}
    \centering
    \includegraphics[width=1\linewidth]{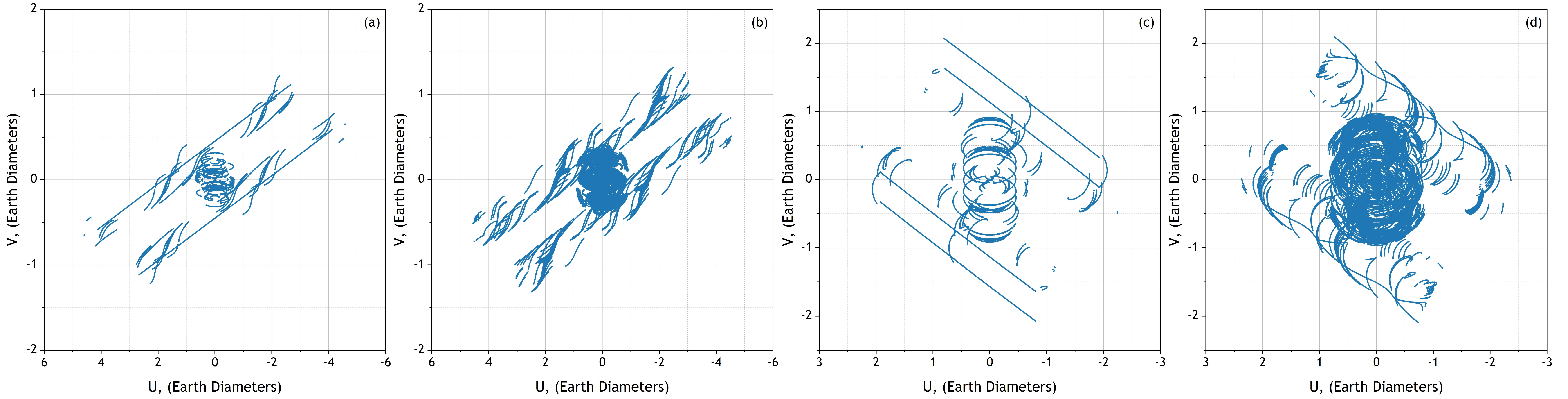}
    \caption{$(u,v)$ coverage for: (a) -- M87 observations together with the EHT telescopes, (b) -- M87 observations together with the ngEHT telescopes, (c) -- Sgr~A* observations together with the EHT telescopes, (d) -- Sgr~A* observations together with ngEHT telescopes. The scale is presented in Earth diameters. The observation frequency is 230~GHz. Duration of observations in for all cases was 48 hours.}
    \label{fig:uv}
\end{figure*}

Fig.~\ref{fig:uv} shows the calculated $(u,v)$ coverage for M87 (a,b) and Sgr~A* (c, d) for EHT (a, c) and next-generation EHT (ngEHT) ground-based telescopes (b, d). The observations lasted 48 hours.

\begin{figure*}
    \centering
   \includegraphics[width=0.65\linewidth]{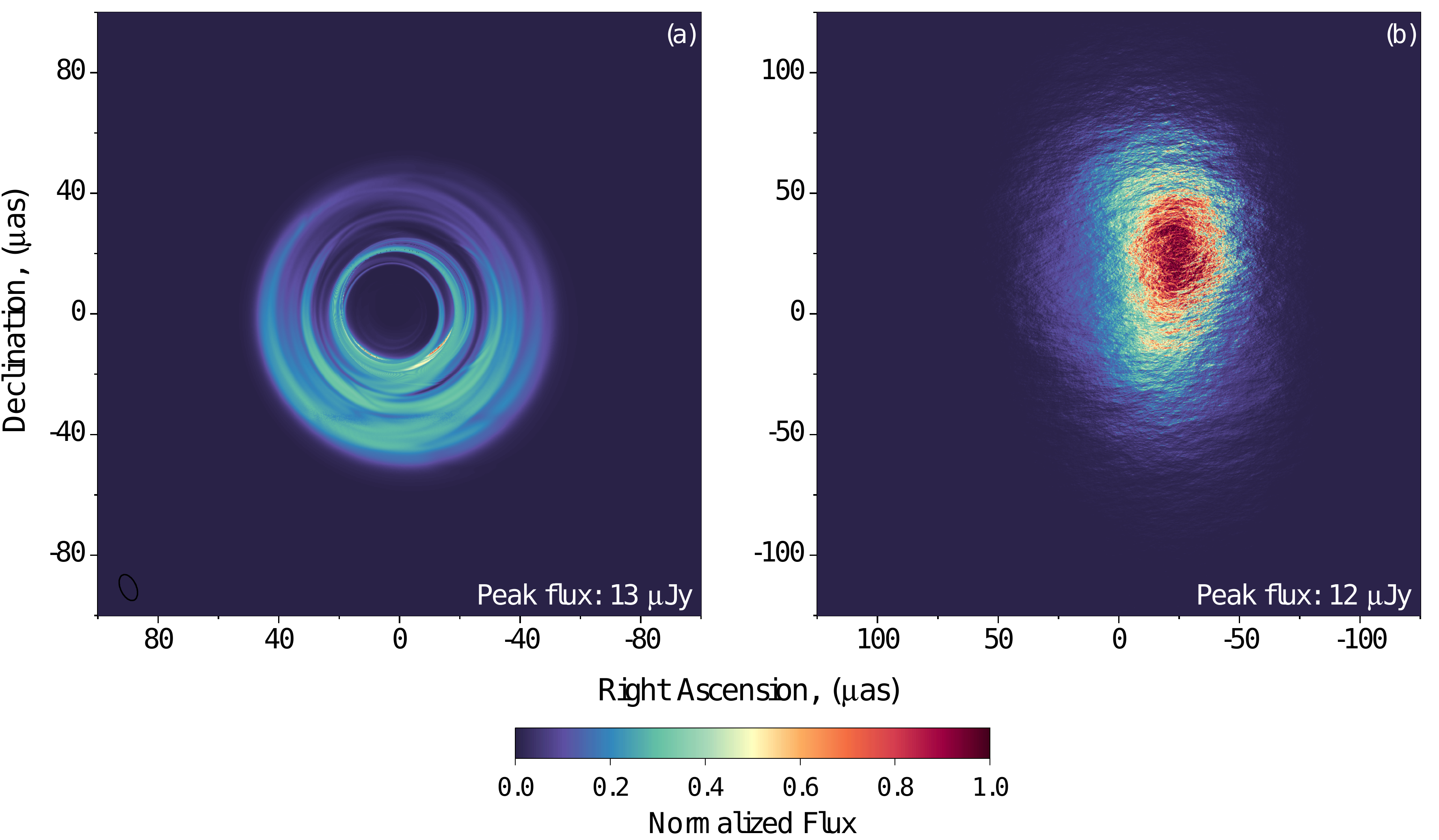}
    \caption{Averaged models at 230~GHz of: (a) -- M87; (b) -- Sgr~A* (scattered). Models described in \cite{Chernov2021}. The colorscale of flux value is normalized to maximum for each image. Scattering parameters for Sgr~A* are taken from \cite{Johnson2019}.}
    \label{fig:mdls}
\end{figure*}

\begin{figure*}
    \centering
    \includegraphics[width=1\linewidth]{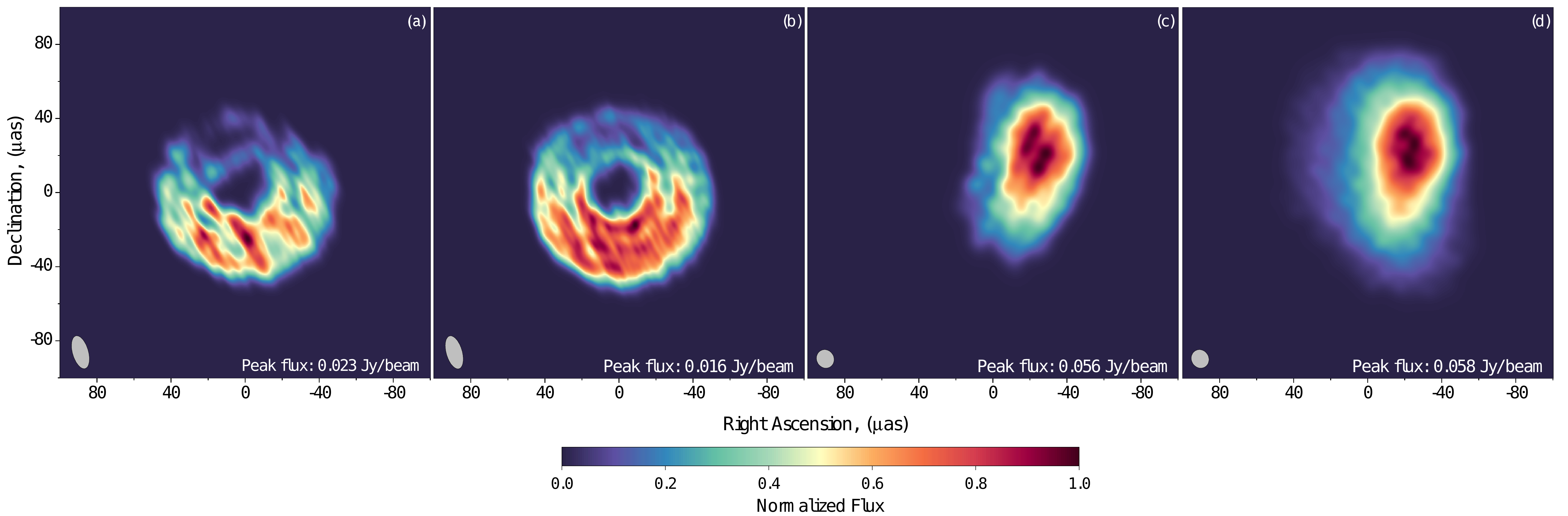}
    \caption{Synthesized averaged images at 230~GHz: (a) -- M87 together with the EHT telescopes, (b) -- M87 together with the ngEHT telescopes, (c) -- Sgr~A* together with the EHT telescopes, (d) -- Sgr~A* together with the ngEHT telescopes. $(u,v)$ coverage is shown in Fig.~\ref{fig:uv} (a)-(d) respectively. Duration of observations in all cases is 48 hours. The colorscale of flux value is normalized to maximum for each image. Grey circles correspond to synthesized beams.}
    \label{fig:imgs}
\end{figure*}

Simulations and analysis were performed with the Astro Space Locator software package \cite{Likhachev2020}. The visibility models of M87 and Sgr~A* were applied to the calculated $(u,v)$ coverage \cite{Chernov2021}. The Sgr~A* model was scattered as described in \cite{Johnson2019}. Fig.~\ref{fig:mdls} shows the M87 (a) and Sgr~A* (b) models, respectively. Then, thermal noise was accounted for by superimposing telescope sensitivity on the simulated visibility amplitude from the model. Sensitivity $SEFD$ expressed in $Jy$ was taken from \cite{EHT2019p2,Likhachev2022} for EHT ground telescopes and from \cite{Raymond2021} for ngEHT telescopes. The estimated $SEFD$ of Millimetron at 230~GHz is $SEFD_{MM}=4000$ Jy \cite{Andrianov2021,Likhachev2022}. A vector with an amplitude of sensitivity (calculated using $SEFD$ values, bandwidth $\Delta\nu=4$~GHz and integration time $t=10$~s) was added to each value of the visibility function of simulated observations. Finally, the images were reconstructed using the classical CLEAN method with a field of view of $0.2\times0.2$~mas, a grid size of $1024\times1024$ cells, and uniform gridding. Fig.~\ref{fig:imgs} shows the average synthesized images of M87 (a,b) and Sgr~A* (c,d) obtained from simulations for these sources.   

\begin{figure*}
    \centering
    \includegraphics[width=0.8\linewidth]{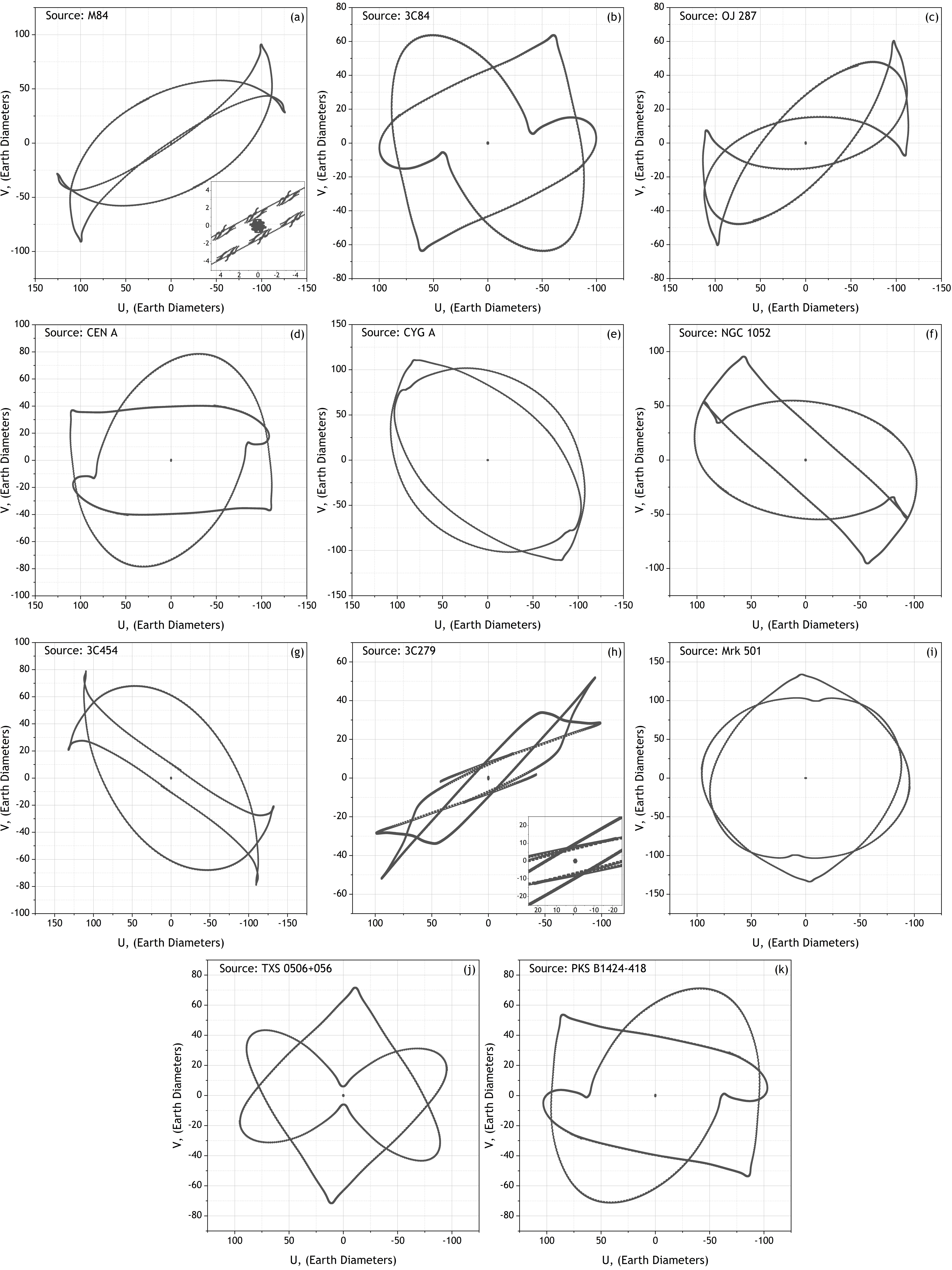}
    \caption{Annual $(u,v)$ coverage for target sources from the Table~\ref{tab:target_sources}, excluding M87 and Sgr~A*: (a) -- M84, (b) -- 3C84, (c) -- OJ287, (d) -- Cen A, (e) -- Cyg A, (f) -- NGC 1052, (g) -- 3C454, (h) -- 3C279, (i) -- Mrk 501, (j) -- TXS 0506+056, (k) -- PKS B1424-418. The scale is presented in Earth diameters. The central region of the $(u,v)$ coverage for M84 is shown separately.}
    \label{fig:uv_coverages}
\end{figure*}

As can be seen from the figures, the resulting orbital configuration allows VLBI observations to obtain images of priority sources M87 and Sgr~A*. As part of the Millimetron space observatory, simultaneous observations at several frequencies are planned. Such a mode has a number of advantages which can significantly increase VLBI observations' scientific value. The first advantage is the improvement of $(u,v)$ coverage, which is critical for space-ground VLBI observations. The second advantage is the ability to compensate ground-based telescopes for phase instabilities caused by atmosphere influence by compensating for phase variations at high frequencies of 230~GHz and above, where the influence of the atmosphere increases significantly, by simultaneous observations at low frequencies \cite{Rioja2017,Rioja2023}. Finally, the $(u,v)$ coverage was calculated for other sources from the Table~\ref{tab:target_sources}. Fig.~\ref{fig:uv_coverages} shows annual $(u,v)$~coverage for sources indicated in the Table~\ref{tab:target_sources}, excluding M87 and Sgr~A*. 

In addition to the two priority sources Sgr~A* and M87, $(u,v)$ coverage for the M84 (Fig.~\ref{fig:uv_coverages} (a)) and 3C279 (Fig.~\ref{fig:uv_coverages} (h)) sources also allows VLBI imaging observations of this source. In general, returning to Fig.~\ref{fig:mol} any source that could fall onto the bold blue line of the satellite trajectory will have relatively good $(u,v)$ coverage potentially suitable for VLBI imaging observations.

\subsection{Radio Visibility}
In addition, it is critical to estimate spacecraft visibility from Earth. This is essential to maintain radio communication and control, measure orbital parameters, and organize data downlink. The observatory's radio visibility from ground stations (GST) was calculated for 10 years. The calculation involved 4 main possible stations for tracking the Millimetron spacecraft and measuring its orbital parameters: Baikonur, Ussuriysk, Bear Lakes, and Evpatoria. For calculations, a topocentric coordinate system based on the WGS-84 geodetic system was used. Restrictions were imposed due to Millimetron's overlap with the Moon. An observatory was considered unobservable if the difference between the directions to it and to the center of the Moon was less than 1.1 times the radius of the Moon's disk.

\begin{figure*}
    \centering
    \includegraphics[width=0.65\linewidth]{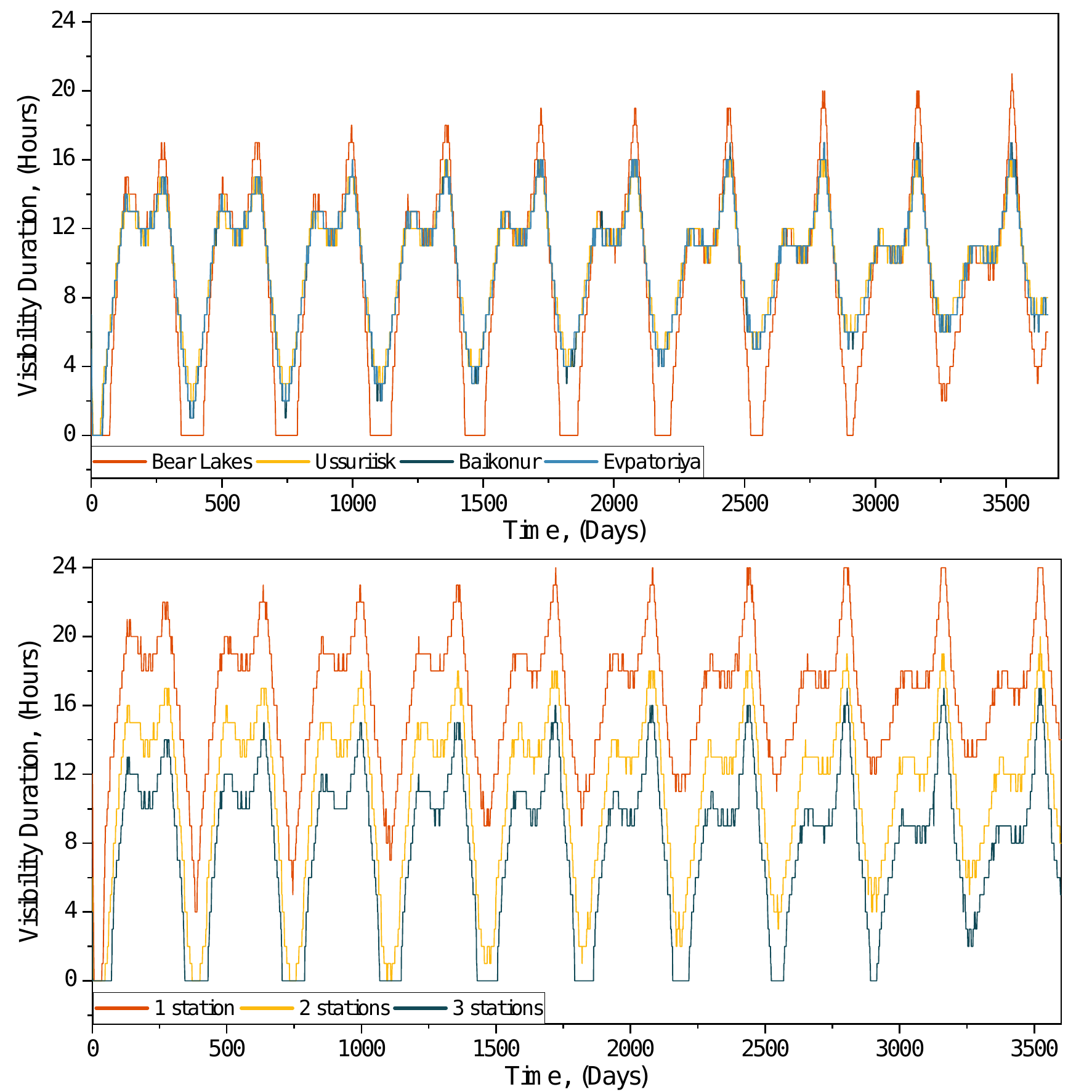}
    \caption{Top panel: daily radio visibility for 10 years of the Millimetron observatory from ground tracking stations: Baikonur, Bear Lakes, Ussuriisk, Evpatoriya. Bottom panel: daily radio visibility when Millimetron observatory is visible by at least one, two or three ground stations simultaneously.}
    \label{fig:vis}
\end{figure*}

Fig.~\ref{fig:vis} shows a graph of radio visibility versus time. For each day of the mission, the number of hours during which the space observatory was visible to the instrument was calculated. As can be seen Bear Lakes have the poorest daily radio visibility zones.

VLBI has another feature. It is related to the fact that VLBI observations can generate huge data flows. In the Millimetron space observatory a bandwidth of $\Delta\nu=4$~GHz in VLBI mode yields a data rate of 8~Gbps with two-bit sampling. Doubling the bandwidth will linearly increase the data rate to 16~Gbps. Having two circular polarizations leads to increased data rates up to 32~Gbps. It is impossible to transmit such data flows in real-time from the L2 point as the transmitting bandwidth is expected to be $\approx1.2~Gbit/s$. Therefore, Millimetron will have onboard data storage of 384~TB. Such an amount of onboard memory allows to record 48000~seconds of VLBI observations conducted in a single frequency range with two side-bands of $\Delta\nu=8$~GHz each and two circular polarizations. This corresponds to a two-day imaging session with scans of 10 seconds and a gap between two scans of 26~seconds. Observation planning in this manner meets the requirements of imaging sessions simulated in Section~\ref{sec:VLBIsim}.

Assuming continuous data transmission, a volume of 384~TB will be transmitted for about 30 days. At the same time, estimates show that radio visibility does not always reach 24 hours per day. However, it is possible to transmit data much longer, since the repeatability of such imaging VLBI observations is determined by the repeatability of the $(u,v)$ coverage for a specific source. Due to the halo orbit's peculiarity, such repeatability is over a year. Estimates show that the average daily radio visibility for 10 years per season is 13 (autumn), 20 (spring), 16 (summer) and 18 (winter) hours. These estimates are satisfactory for Millimetron observatory and provide a successful transmission of 384~TB within 40-50 days.

\section{Results and Conclusions}
A variant of the Millimetron observatory orbit was obtained. For further orbit design an analytical halo orbit in the CR3BP system with an altitude of $3.7\times10^5$~km above the ecliptic plane was used. This orbit was transferred to a real force model and analyzed by date to ensure scientific program implementation. Unlike previous works that used analytical methods, this work employs numerical integration for orbital design.

The exact dates for departure, halo formation, and orbital correction were calculated. Start time of the first orbit: 2030-09-11T01:50:02.6. The flight to a halo orbit is carried out in an intermediate circular orbit around the Earth at 200~km. The launch time from the intermediate orbit is 2030-05-18T22:31:53.5. Time of arrival at the L2 point orbit is 2030-09-25T00:40:02.6. To form a halo orbit, an impulse of 7.87~m/s ($\approx$19~kg of fuel) is required. Maintenance corrections for this orbit over 10 years do not exceed 11~m/s. 

The estimated daily radio visibility using four ground control stations for the resulting halo orbit will be at least 4 hours when the satellite is visible from at least one of the stations. Overall this is a satisfactory result. However, at least it can be concluded that it is necessary to use ground tracking stations in the southern hemisphere to increase the radio visibility time of the Millimetron observatory.

Scientific program feasibility was assessed for two observatory operating modes. In single telescope mode, 98\% of the celestial sphere is covered within a year. Based on the orbital design methods developed in this work, we investigated the existence of the shortest baseline projection and its specific dates for the space-ground VLBI mode. It is shown that the calculated orbit satisfies the requirements for imaging observations of two priority sources: M87 and Sgr~A*. Optimal baseline projections for M87 are achieved on 2031-03-15 with a minimum baseline projection of 11637~km, for Sgr~A* on 2031-07-20 and 10827~km, respectively. The feasibility of scientific objectives was demonstrated through VLBI visibility simulation and image reconstruction.

Moreover, it is also possible to image M84, 3C279 and other objects that fit on the observatory trajectory in the Molweide projection. To improve the quality of $(u,v)$ coverage and increase the probability of signal detection during correlation processing of VLBI observations, it is planned to use multi-frequency synthesis.

To summarize, the resulting halo orbit fully meets all observatory technical requirements. The assessments performed for two observation modes showed that the orbit also meets the requirements for organizing scientific observations both in the single-dish and VLBI modes. Separately, it should be noted that the Millimetron observatory will be the first observatory in the world that will conduct space-ground interferometric observations in a halo orbit at a distance of 1.5 million~km from Earth.

\section*{Data availability}
The data underlying this article and the results of the simulations will be shared on reasonable request to the corresponding author.

\bibliographystyle{elsarticle-num}
\bibliography{biblio}
\end{document}